\documentclass[aps]{revtex4}%
\usepackage{amsmath}
\usepackage{graphicx}%

\usepackage{amsfonts}%
\usepackage{amssymb}

\begin{document}
\title{Phenomenological description of quantum gravity inspired modified classical
electrodynamics \footnote{Dedicated to Octavio Obreg\'{o}n on his
sixthiest birthday} }
\author{R. Montemayor$^1$ and Luis F. Urrutia$^2$}
\affiliation{$^1$ Instituto Balseiro and CAB, Universidad Nacional
de Cuyo and CNEA, 8400
Bariloche, Argentina\\
$^2$ Instituto de Ciencias Nucleares, Universidad Nacional Aut{\'o}noma de M{%
\'e}xico, A. Postal 70-543, 04510 M{\'e}xico D.F., M{\'e}xico}

\begin{abstract}
We discuss a large class of phenomenological models incorporating
quantum gravity motivated corrections to electrodynamics. The
framework is that of electrodynamics in a birefringent and
dispersive medium with non-local constitutive relations, which are
considered up to second order in the inverse of the energy
characterizing the quantum gravity scale. The energy-momentum
tensor, Green functions and frequency dependent refraction indices
are obtained,  leading to departures from standard physics. The
effective character of the theory is also emphasized by
introducing a frequency cutoff $\Omega$. The analysis of its
effects upon the standard notion of causality is performed,
showing that in the radiation regime ($\Omega R>> 1$) the expected
corrections of the order $\left({\omega}/{\Omega}\right)^n$ get
further suppressed by highly oscillating terms proportional to
$\sin(\Omega R), \ \cos(\Omega R)$, thus  forbiding causality
violations to show up in the corresponding observational effects.

\end{abstract}
\maketitle

\baselineskip=20pt

\section{Introduction}

Lorentz covariance in local inertial frames is a well established
symmetry at the energies of present day experiments. However, its
validity at high energies is subject to test. Possible Lorentz
invariance violations may arise from dynamical modifications
induced by quantum gravity (QG). The effects of such violations in
the range well below the Planck energy $(E_{P}\sim 10^{19}\,GeV)$
have been recently the object of intense scrutiny
\cite{GAC,GPED,URRU,analysis,JACOBSON}. This issue is closely
linked to theoretical and
experimental research, based on the Standard Model Extension
\cite{KOSTELECKY}%
, concerning Lorentz and CPT violations \cite{CPTMEETS}. Heuristic
loop QG derivations of such effects \cite{GPED,URRU} make it clear
that a better understanding of the corresponding semiclassical
limit is required \cite{THIEMANN}. String theory has also provided
models for explaining such QG induced corrections \cite{QFMODEL}.
Moreover, effective field theory models have been constructed that
include higher dimension Lorentz invariance violating (LIV)
operators \cite{MP}. Synchrotron radiation arising from the model
in \cite{MP} has been extensively analyzed in Ref. \cite{MU}.
These effective theories use a reduced number of degrees of
freedom to describe the physics at a low energy scale, ignoring
the detailed dynamics inherent to Planck energies. In other words,
if QG dominates at a scale $E_{QG}$, usually considered of the
order of $E_{P}$, a corresponding low energy effective theory can
be visualized as an expansion in powers of $\,\;\tilde{\xi}\simeq
E_{QG}^{-1}$, truncated at a given finite order. In this way it
will be a good description, hopefully simpler than the original
one, for energies $E\ll E_{QG}$. This restricted validity relaxes
some of the constraints usually required for physical theories,
such as renormalizability. Stability and causality, perhaps of
more essential status than the Lorentz symmetry itself, are
assumed to remain valid at the low energy regime \cite{LEHNERT}.
Nevertheless, fine tuning problems arise when considering
radiative corrections \cite{CPSUV}, which can be circumvented by
extending the notion of dimensional regularization \cite{ALFARO}.

In fact, one of the possible manifestations of QG at low energy is
the appearance of correction terms related to the scale $E_{QG}$\
in the standard particle propagation and interaction properties.
The most direct interpretation of such corrections, though not the
only one \cite{ADDINT}, is in terms of a spontaneous breaking of
Lorentz covariance at high energies. If this is so, the effective
theory will be covariant under Lorentz transformation between
inertial frames (passive or observer transformations), and the
observable Lorentz symmetry violations will be associated with
rotations and boosts of the fields in a given inertial frame
(active or particle transformations). In this case the space-time
coordinates at low energy remain commutative. We focuse here on a
general analysis of QG effects in electrodynamics in these models,
where we can introduce the full usual mathematical framework of
field theory, specially Fourier transformations. QG may also
modify the space-time itself so that the coordinates become
noncommutative, as in the case of  Double Special Relativity
models, for example. In this situation, which will not be
discussed here, ordering ambiguities preclude a direct use of such
transformations.

Considering now the electromagnetic field, the models proposed to describe low
energy effects of QG can usually be expressed in terms of modified dispersion
relations, with a polynomial dependence in energy and momentum. Such
modifications include standard Lorentz invariance violations as well as
possible extensions of Lorentz covariance \cite{REVIEWS}. Most of these
approaches can be unified in the description%
\begin{align}
&  i\mathbf{k}\cdot\mathbf{D}=4\pi\rho,\;\;\;\;\;\;\mathbf{k}\cdot
\mathbf{B}=0,\label{GENMAXW1}\\
&  \mathbf{k}\times\mathbf{E}-\omega\mathbf{B}=0,\;\;i\mathbf{k}%
\times\mathbf{H}+i\omega\mathbf{D}=4\pi\mathbf{j},\label{GENMAXW2}%
\end{align}
where the auxiliary fields%
\begin{equation}
D^{i}=\alpha^{ij}E_{j}+\rho^{ij}B_{j},\quad H^{i}=\beta^{ij}B_{j}+\sigma
^{ij}E_{j},\label{CRH}%
\end{equation}
are such that the coefficients $\alpha^{ij}$, $\beta^{ij}$,
$\rho^{ij}$ and $\sigma^{ij}$ depend on the energy $\omega$ and
the momentum $\mathbf{k}$ of the electromagnetic field. These
equations correspond to a higher order linear dynamics. Equations\
(\ref{GENMAXW1}) and (\ref{GENMAXW2}) strongly resemble the usual
description for an electromagnetic field in a medium
\cite{LANDAU}, where the fields $\mathbf{D}$ and $\mathbf{H}$ are
characterized by constitutive relations of the form (\ref{CRH}).
In terms of electrodynamics in media, we can interpret the low
energy QG corrections in terms of a dispersive bianisotropic
media. From a heuristic point of view, as shown below, these
effective media are non-local in space and time, which can be
interpreted as a footprint of the granularity induced by QG.

Strictly speaking, the effective models are characterized by
Eqs.(\ref{GENMAXW1}-\ref{GENMAXW2}), but under certain restrictions it is also
possible to pose an action from which they derive. Although not essential,
this is a useful approach to visualize general features of the dynamics. Let
us recall the Lagrangian for the electromagnetic field in a local medium%
\begin{equation}
L=-\frac{1}{4}F_{\mu\nu}\chi^{\left[  \mu\nu\right]  \left[  \alpha
\beta\right]  }F_{\alpha\beta}-4\pi j_{\mu}A^{\mu},\label{L}%
\end{equation}
where $F_{\mu\nu}=\partial_{\mu}A_{\nu}-\partial_{\nu}A_{\mu}$, and
$\chi^{\left[  \mu\nu\right]  \left[  \alpha\beta\right]  }$ contains the
information about the medium. This structure warrants gauge invariance and
hence charge conservation. The dynamics is given by the equations of motion%
\begin{equation}
\partial_{\mu}H^{\mu\nu}=4\pi j^{\nu},\label{cem}%
\end{equation}
together with the constitutive relations%
\begin{equation}
H^{\mu\nu}=\chi^{\left[  \mu\nu\right]  \left[  \alpha\beta\right]  }%
F_{\alpha\beta}.\label{ccr}%
\end{equation}
Defining the electric and magnetic fields as $F_{0i}=E_{i}\;$and
$F_{ij}=-\epsilon_{ijk}B_{k}$ respectively, and the corresponding components
of $H^{\mu\nu}$, $H^{0i}=D^{i}$ and $H^{ij}=-\epsilon^{ijk}H^{k}$, the
constitutive relations become%
\begin{equation}
D^{i}=2\chi^{\left[  0i\right]  \left[  0j\right]  }E_{j}-\chi^{\left[
0i\right]  \left[  mn\right]  }\epsilon_{mnj}B_{j},\quad H^{i}=\epsilon
_{ilk}\chi^{\left[  lk\right]  \left[  0j\right]  }E_{j}-\frac{1}{2}%
\epsilon_{ilk}\chi^{\left[  lk\right]  \left[  mn\right]  }\epsilon_{mnj}%
B_{j}.\label{CR2}%
\end{equation}
In our notation the components of any three-dimensional vector $\mathbf{V\;}%
$are given by those with subindices $V_{i}$.

The equations of motion incorporating QG corrections acquire a similar form,
but with one important difference arising from the nonlocal character of the
effective medium. The $\chi^{\left[  \mu\nu\right]  \left[  \alpha
\beta\right]  }$ is now a non-local tensor, such that%
\begin{equation}
L=-\frac{1}{4}\int\;d^{4}\tilde{x}\;F_{\mu\nu}(x^{\sigma})\chi^{\left[  \mu
\nu\right]  \left[  \alpha\beta\right]  }(x^{\sigma}-\tilde{x}^{\sigma
})\;F_{\alpha\beta}(\tilde{x}^{\sigma})-4\pi j_{\mu}(x^{\sigma})A^{\mu
}(x^{\sigma}),\label{lt}%
\end{equation}
instead of (\ref{L}). As usual if $F_{\mu\nu}$ and $L$ are real, the reality
of $\chi^{\left[  \mu\nu\right]  \left[  \alpha\beta\right]  }(x^{\sigma
}-\tilde{x}^{\sigma})$ and subsequently of
\begin{equation}
H^{\mu\nu}(x)=\int\;d^{4}\tilde{x}\chi^{\left[  \mu\nu\right]  \left[
\alpha\beta\right]  }(x^{\sigma}-\tilde{x}^{\sigma})\;F_{\alpha\beta}%
(\tilde{x}^{\sigma}).
\end{equation}
are implied. Writing $\chi^{\left[  \mu\nu\right]  \left[  \alpha\beta\right]
}(x^{\sigma}-\tilde{x}^{\sigma})$ in terms of its Fourier transform%
\begin{equation}
\chi^{\left[  \mu\nu\right]  \left[  \alpha\beta\right]  }(x^{\sigma}%
-\tilde{x}^{\sigma})=\int d^{4}k\;e^{-ik\cdot(x-\tilde{x})}\;\chi^{\left[
\mu\nu\right]  \left[  \alpha\beta\right]  }(k^{\sigma}),
\end{equation}
we can easily demonstrate that (\ref{lt}) can also be written a
\begin{equation}
L=-\frac{1}{4}F_{\mu\nu}(x^{\sigma})\;\left[  {\hat{\chi}}^{\left[  \mu
\nu\right]  \left[  \alpha\beta\right]  }(i\partial_{\sigma})\;F_{\alpha\beta
}(x^{\sigma})\right]  ,\label{LNL}%
\end{equation}
with $\hat{\chi}^{\left[  \mu\nu\right]  \left[  \alpha\beta\right]  }%
$\textbf{\ }being a derivative operator. In terms of the Fourier transform,
the reality of $\chi^{\left[  \mu\nu\right]  \left[  \alpha\beta\right]
}(x^{\sigma}-\tilde{x}^{\sigma})$\ is stated as%
\begin{equation}
\left[  \chi^{\left[  \mu\nu\right]  \left[  \alpha\beta\right]  }(k^{\sigma
})\right]  ^{\ast}=\chi^{\left[  \mu\nu\right]  \left[  \alpha\beta\right]
}(-k^{\sigma}),
\end{equation}
which holds similarly for the transformed fields$\;F_{\alpha\beta}(k^{\sigma
})$\ and $H_{\alpha\beta}(k^{\sigma})$.$\;$If $\hat{\chi}^{\left[  \mu
\nu\right]  \left[  \alpha\beta\right]  }$ is symmetric, in the sense that for
each set of index values ($\mu$,$\nu$,$\alpha$,$\beta$) (no sum with respect
to repeated indices)%
\begin{equation}
\int d^{4}x\;F_{\mu\nu}\left(  {\hat{\chi}}^{\left[  \mu\nu\right]  \left[
\alpha\beta\right]  }F_{\alpha\beta}\right)  =\int d^{4}x\;F_{\alpha\beta
}\left(  {\hat{\chi}}^{\left[  \alpha\beta\right]  \left[  \mu\nu\right]
}F_{\mu\nu}\right)  \label{sym}%
\end{equation}
is satisfied, then it is possible to perform integrations by parts making the
equations of motion of the same form as in the usual non-operator case. Thus
the Fourier transform of the equations of motion and constitutive relations
acquire the structure of (\ref{GENMAXW1}-\ref{GENMAXW2}) and (\ref{CRH})
respectively. In the following we assume that this property is satisfied. When
the components of $\hat{\chi}^{\left[  \mu\nu\right]  \left[  \alpha
\beta\right]  }$ do not correspond to a standard Lorentz tensor, this
Lagrangian describes a model where the Lorentz symmetry is broken by the medium.

We use this approach, where the QG modifications are described
phenomenologically by constitutive relations, to discuss the main properties
of QG induced effects in electrodynamics. A low energy expansion is developed
in terms of the parameter ${\tilde{\xi}}$ $\simeq E_{QG}^{-1}$. Working to
order ${\tilde{\xi}}^{2}$ allows us to present $\hat{\chi}^{[\mu\nu
][\alpha\beta]}\;$in the form%
\begin{equation}
\hat{\chi}^{[\mu\nu][\alpha\beta]}=\chi_{0}^{[\mu\nu][\alpha\beta]}+\chi
_{1}^{[\mu\nu]\theta\lbrack\alpha\beta]}\partial_{\theta}+\chi_{2}^{[\mu
\nu]\{\theta\psi\}[\alpha\beta]}\partial_{\theta}\partial_{\psi}%
,\label{GENSUS}%
\end{equation}
where the constant coefficients $\chi_{1}^{[\mu\nu]\theta\lbrack\alpha\beta
]},\chi_{2}^{[\mu\nu]\{\theta\psi\}[\alpha\beta]}$ are proportional to
$\tilde{\xi}\;,\tilde{\xi}^{2}\;$respectively. They are antisymmetric in the
indices inside square brackets and symmetric in the indices inside curly
brackets. In this way we are considering a Lagrangian depending up to third
derivatives in the basic electromagnetic potential $A_{\mu}.\;$Moreover, the
integration conditions (\ref{sym}) require the following symmetry properties%
\begin{equation}
\chi_{0}^{[\mu\nu][\alpha\beta]}=\chi_{0}^{[\alpha\beta][\mu\nu]},\;\chi
_{1}^{[\mu\nu]\theta\lbrack\alpha\beta]}=-\;\chi_{1}^{[\alpha\beta
]\theta\lbrack\mu\nu]}\;,\;\;\chi_{2}^{[\mu\nu]\{\theta\psi\}[\alpha\beta
]}=\chi_{2}^{[\alpha\beta]\{\theta\psi\}[\mu\nu]}.
\end{equation}
Once the coefficients of the constitutive relations have been promoted to
derivative operators\ we obtain the relations
\begin{align}
{\hat{\chi}}^{\left[  0i\right]  \left[  0j\right]  }  & =\frac{1}{2}%
\hat{\alpha}^{ij},\ \ \ \ \ \ \ \ \ \ \ \ \ \ \ \ \ \ \ {\hat{\chi}}^{\left[
0i\right]  \left[  mn\right]  }=-\frac{1}{2}\epsilon_{mnj}\hat{\rho}^{ij},\\
{\hat{\chi}}^{\left[  mn\right]  \left[  0j\right]  }  & =+\frac{1}{2}%
\epsilon_{mni}\hat{\sigma}^{ij},\ \ \ \ \ \ \ \ \ {\hat{\chi}}^{\left[
lk\right]  \left[  mn\right]  }=+\frac{1}{2}\epsilon_{kli}\hat{\beta}%
^{ij}\epsilon_{jmn},
\end{align}
by comparing Eqs. (\ref{CR2}) and (\ref{CRH}). The expansion\
(\ref{GENSUS}) induces the corresponding form in the coefficients
of the constitutive relations
\begin{equation}
\hat{\alpha}^{ij}=\alpha_{0}^{\left(  ij\right)  }+\tilde{\xi}\alpha
_{1}^{\left(  ij\right)  \psi}\partial_{\psi}+\tilde{\xi}^{2}\alpha
_{2}^{\left(  ij\right)  \psi\theta}\partial_{\psi}\partial_{\theta
},\label{ALPHA}%
\end{equation}
and similarly for $\hat{\rho}^{ij},\, \hat{\sigma}^{ij}$ and $\hat{\beta}%
^{ij}$, in an obvious notation. Here $\alpha_{0}^{\left(  ij\right)  }%
,\alpha_{1}^{\left(  ij\right)  \psi}$and $\alpha_{2}^{\left(  ij\right)
\psi\theta}$ are constant coefficients.

To analyze the propagation of the fields and to define the
corresponding refraction index, the dependence of the constitutive
relations on $\omega$ and $\mathbf{k}$ has to be made explicit. To
achieve this we first expand the coefficients of the constitutive
relations in space derivatives, maintaining covariance under
rotations. Considering that these\textbf{\ }models can be
understood as perturbative descriptions in terms of the parameter
${\tilde {\xi}}$ we have, up to order ${\tilde{\xi}}^{2}$,
\begin{equation}
\hat{\alpha}^{ij}=\alpha_{0}(\partial_{t})\eta^{ij}+\alpha_{1}(\partial
_{t}){\tilde{\xi}}\epsilon^{ijr}\partial_{r}+\alpha_{2}(\partial_{t}%
){\tilde{\xi}}^{2}\partial^{i}\partial^{j},\label{alpha}%
\end{equation}
with analogous expansions for $\hat{\beta}^{ij},\hat{\sigma}^{ij}$ and
$\hat{\rho}^{ij}$. Here $\alpha_{A}$, $\beta_{A}$, $\sigma_{A}$, and $\rho
_{A}$, with $A=0,1,2$, are $S0(3)$ scalar operators. This approach can be
generalized to models with preferred spatial directions. The symmetry of
${\hat{\chi}}$ in Eq. (\ref{sym}) implies that the terms in $\hat{\alpha}%
^{ij},$ $\hat{\rho}^{ij}$and $\hat{\beta}^{ij}$ with an even number of
derivatives are symmetric under $i\leftrightarrow j$, while the terms with an
odd number of derivatives are antisymmetric. In the case of $\rho^{ij}$ and
$\sigma^{ij}$ Eq. (\ref{sym}) leads to%
\begin{equation}
\rho_{A}=-\sigma_{A}.
\end{equation}
Furthermore, we can also consistently expand the coefficients $\alpha_{A}$,
$\beta_{A}$, $\sigma_{A}$, and $\rho_{A}$ in powers of ${\tilde{\xi}\;}$,
according to
\begin{equation}
\zeta_{A}\simeq\zeta_{A0}+\zeta_{A1}{\tilde{\xi}}\partial_{t}+\zeta
_{A2}{\tilde{\xi}}^{2}\partial_{t}^{2},\;\;\;\zeta_{A}=\{\alpha_{A},\beta
_{A},\rho_{A},\sigma_{A}\},\;\;\;
\end{equation}
with $\zeta_{A0},$ $\zeta_{A1},\;\zeta_{A2}\;$being constant coefficients.
This expansion leads to the identification of the following coefficients
\begin{align}
\alpha_{0}^{\left(  ij\right)  }  & =\beta_{0}^{\left(  ij\right)  }=\eta
^{ij},\;\;\rho_{0}^{\left(  ij\right)  }=0,\;\;\alpha_{1}^{\left(  ij\right)
0}=\beta_{1}^{\left(  ij\right)  0}=\rho_{1}^{\left(  ij\right)
0}=0,\nonumber\\
\alpha_{1}^{\left(  ij\right)  r}  & =\epsilon^{ijr}\alpha_{10},\;\beta
_{1}^{\left(  ij\right)  r}=\epsilon^{ijr}\beta_{10},\;\rho_{1}^{\left(
ij\right)  r}=\epsilon^{ijr}\rho_{10},\;\;\;\;\;\nonumber\\
\alpha_{2}^{\left(  ij\right)  00}  & =\eta^{ij}\alpha_{00},\;\;\beta
_{2}^{\left(  ij\right)  00}=\eta^{ij}\beta_{02}^{2},\;\rho_{2}^{\left(
ij\right)  00}=\eta^{ij}\rho_{02},\;\;\alpha_{2}^{\left(  ij\right)  0p}%
=\beta_{2}^{\left(  ij\right)  0p}=\rho_{2}^{\left(  ij\right)  0p}%
=0,\nonumber\\
\alpha_{2}^{\left(  ij\right)  mn}  & =\frac{1}{2}\left(  \delta^{im}%
\delta^{jn}+\delta^{in}\delta^{jm}\right)  \alpha_{20},\;\;\beta_{2}^{\left(
ij\right)  mn}=\frac{1}{2}\left(  \delta^{im}\delta^{jn}+\delta^{in}%
\delta^{jm}\right)  \beta_{20},\ \ \rho_{2}^{\left(  ij\right)  mn}=\frac
{1}{2}\left(  \delta^{im}\delta^{jn}+\delta^{in}\delta^{jm}\right)  \rho
_{20}.\label{CONSTCOEF}%
\end{align}
The above partition is consistent with the requirement of covariance under
rotations. We have taken $\alpha_{00}=\beta_{00}=1$ and $\rho_{00}=\sigma
_{00}=0$ to recover the usual vacuum as the background for ${\tilde{\xi}=0}$.

The fact that this theory does not hold at high energies will be coded by
cutoff $\Omega\ll E_{QG}$. We provide a general description of such modified
electrodynamics including expressions for the equations of motion, the
energy-momentum tensor\ and the Green functions as well as the corresponding
refraction indexes, up to second order in $\tilde{\xi}$.

\section{Equations of motion}

Equations (\ref{alpha}), together with the corresponding ones for the
remaining coefficients of the constitutive relations give%
\begin{align}
\mathbf{D} &  =\left(  \alpha_{0}+\alpha_{2}{\tilde{\xi}}^{2}\mathbf{k}%
^{2}\right)  \mathbf{E}-\left(  \sigma_{0}+i\alpha_{1}{\tilde{\xi}\omega
}\right)  \mathbf{B}+\left(  i\sigma_{1}{\tilde{\xi}}+\omega\alpha_{2}%
{\tilde{\xi}}^{2}\right)  \left(  \mathbf{k\times B}\right)  ,\label{D1}\\
\mathbf{H} &  =\left(  \beta_{0}-i{\omega}\sigma_{1}{\tilde{\xi}}\right)
\mathbf{B}-i\beta_{1}{\tilde{\xi}}\left(  \mathbf{k\times B}\right)
+\sigma_{0}\mathbf{E}.\label{H1}%
\end{align}
In the approximation to order ${\tilde{\xi}}^{2}$ here considered, we have
${\tilde{\xi}}^{2}k^{2}\simeq{\tilde{\xi}}^{2}\omega^{2}$ and we can write%
\begin{align}
\mathbf{D} &  =d_{1}(\omega)\mathbf{E}+id_{2}(\omega)\mathbf{B}+d_{3}%
(\omega){\tilde{\xi}}\left(  \mathbf{k\times B}\right)  ,\label{D2}\\
\mathbf{H} &  =h_{1}(\omega)\mathbf{B}+i{h}_{2}(\omega)\mathbf{E}%
+ih_{3}(\omega){\tilde{\xi}}\left(  \mathbf{k\times B}\right)  ,\label{H2}%
\end{align}
where the functions $d_{i}(\omega)$ and $h_{i}(\omega)$ depend only on
$\omega$ and admit a series expansion in powers of ${\tilde{\xi}}\omega$,
characterizing each specific model. From Eqs. (\ref{GENMAXW1}-\ref{GENMAXW2})
we get the equations for $\mathbf{E}$ and $\mathbf{B}$%
\begin{align}
id_{1}\left(  \mathbf{k}\cdot\mathbf{E}\right)   &  =4\pi\rho\left(
\omega,\mathbf{k}\right)  ,\label{INHOM1}\\
i\omega d_{1}\mathbf{E}+\left(  h_{3}k^{2}-g(\omega)\right)  {\tilde{\xi}%
}\mathbf{B}+\left(  \omega d_{3}{\tilde{\xi}+}h_{1}\right)  \left(
i\mathbf{k}\times\mathbf{B}\right)   &  =4\pi\mathbf{j}\left(  \omega
,\mathbf{k}\right)  ,\label{INHOM2}%
\end{align}
where we denote%
\begin{equation}
\left(  d_{2}+h_{2}\right)  \omega=g(\omega){\tilde{\xi}}.\label{INHOM0}%
\end{equation}
The expressions (\ref{D1}-\ref{H1}) indeed indicate that the above
combination\ is of order ${\tilde{\xi}}$. We thus see that in fact there are
only three independent functions of $\omega$ and $k$ which determine the
dynamics%
\begin{equation}
P=d_{1},\quad Q=h_{1}+\omega d_{3}{\tilde{\xi}},\quad R=\left(  h_{3}%
k^{2}-g(\omega)\right)  {\tilde{\xi}}.\label{R}%
\end{equation}
Using the homogeneous equation $\omega\mathbf{B}=\mathbf{k}\times\mathbf{E}$
that yields $\omega\left(  \mathbf{k}\times\mathbf{B}\right)  =\left(
\mathbf{k\cdot E}\right)  \mathbf{k-}k^{2}\mathbf{E}$, and charge conservation
$\omega\rho-\mathbf{k\cdot J}=0$, we decouple the equations for fields
$\mathbf{E}$ and $\mathbf{B}$. Finally, we introduce the standard potentials
$\Phi$ and $\mathbf{A}$
\begin{equation}
\mathbf{B}=i\mathbf{k}\times\mathbf{A},\qquad\mathbf{E}=i\omega\mathbf{A-}%
i\mathbf{k}\Phi.
\end{equation}
in the radiation gauge, $\mathbf{k}\cdot\mathbf{A=}0$, in which case we have
\begin{align}
\Phi &  =4\pi\left(  k^{2}P\right)  ^{-1}\rho,\label{PHIRG}\\
\left(  k^{2}Q-\omega^{2}P\right)  \mathbf{A}+iR\left(  \mathbf{k}%
\times\mathbf{A}\right)   &  =4\pi\left[  \mathbf{j}-\left(  \mathbf{j}%
\cdot\mathbf{\hat{k}}\right)  \mathbf{\hat{k}}\right]  =4\pi\mathbf{j}%
_{T},\label{EQARG}%
\end{align}
from Eqs. (\ref{INHOM1}-\ref{INHOM2}). The presence of birefringence depends
on the parity violating term proportional to\textbf{\ }$R$\textbf{.} It is
clear that a diagonalization is obtained in a circular polarization basis.
Decomposing the vector potential and the current in such a basis%
\begin{equation}
\mathbf{A}=\mathbf{A}^{+}+\mathbf{A}^{-},\;\;\;\;\mathbf{j}_{T}=\mathbf{j}%
_{T}^{+}+\mathbf{j}_{T}^{-},
\end{equation}
and recalling the basic properties $\mathbf{\hat{k}}\times\mathbf{A}%
^{+}=-i\mathbf{A}^{+}$, $\mathbf{\hat{k}}\times\mathbf{A}^{-}=i\mathbf{A}^{-}%
$, we separate (\ref{EQARG}) into the decoupled equations \cite{PDELC}
\begin{equation}
\left[  k^{2}Q-\omega^{2}P+\lambda k\,R\right]  \mathbf{A}^{\lambda}%
=4\pi\mathbf{j}_{T}^{\lambda},\;\;\;\;\lambda=\pm1.\label{af}%
\end{equation}
In terms of the basic functions introduced in the constitutive relations
(\ref{D2}-\ref{H2}), the factor in (\ref{af}) is rewritten as
\begin{equation}
k^{2}Q-\omega^{2}P+\lambda k\,R=\lambda h_{3}k^{3}{\tilde{\xi}+}\left(
h_{1}+d_{3}{\omega\tilde{\xi}}\right)  k^{2}-\lambda g{\tilde{\xi}}%
k-d_{1}\omega^{2}.
\end{equation}
This is the key expression to obtain the Green functions and the refraction indices.

\section{Generalized Energy-Momentum tensor}

Any application of this modified electrodynamics related to
radiation and its properties requires the construction of the
corresponding energy momentum tensor. This section is devoted to
such a construction.\textbf{\ }The theories under consideration
are of higher order in the field derivatives, and thus call for an
extension of the standard Noether theorem. The manipulations are
highly simplified by proceeding in a covariant notation, the point
being that the tensorial\textbf{\ }operator\textbf{\
}$\hat{\chi}^{[\mu\nu][\alpha\beta ]}$\textbf{\ }is\textbf{\
}constructed in a given reference frame and satisfies only passive
Lorentz covariance. We assume that active Lorentz invariance is
violated while active translation invariance is maintained, so
that there is an energy momentum tensor given by the Noether
theorem.

Before constructing this tensor in our particular case we recall the general
formalism for\textbf{\ }a Lagrangian including up to three derivatives in the
fields. This can be useful also when consistently including dimension five and
six operators in the matter field coupled to the above electrodynamics. We
start from an action of the form%
\begin{equation}
S=\int d^{4}x\;L(\Phi_{A},\;\Phi_{A,\mu},\;\partial_{\nu}\Phi_{A,\mu
},\;\partial_{\nu}\partial_{\rho}\Phi_{A,\mu}),\label{ACTION3}%
\end{equation}
where we consider $\Phi_{A}$,$\;\Phi_{A,\mu}$,$\;\partial_{\nu}\Phi_{A,\mu}%
$,$\;\partial_{\nu}\partial_{\rho}\Phi_{A,\mu}$, to be independent
fields, i.e.
at this level\textbf{\ }we take for example that\textbf{\ }$\partial_{\nu}%
\Phi_{A,\mu}\neq\partial_{\mu}\Phi_{A,\nu}$. Applying the standard action
principle one derives the Euler-Lagrange (EL)\ equations%
\begin{equation}
0=\frac{\delta L}{\delta\Phi_{A}}-\partial_{\mu}\left(  \frac{\delta L}%
{\delta\Phi_{A,\mu}}\right)  +\partial_{\mu}\partial_{\nu}\left(  \frac{\delta
L}{\delta\left(  \partial_{\nu}\Phi_{A,\mu}\right)  }\right)  -\partial_{\mu
}\partial_{\nu}\partial_{\rho}\left(  \frac{\delta L}{\delta\left(
\partial_{\nu}\partial_{\rho}\Phi_{A,\mu}\right)  }\right)  .\label{GENEL}%
\end{equation}
Assuming that translations generated by $x^{\prime\mu}=x^{\mu}+a^{\mu}$\ are a
symmetry of the action (\ref{ACTION3}), Noether's theorem leads to the energy
momentum tensor%
\begin{align}
T_{\;\;\sigma}^{\tau}  & =-\delta_{\sigma}^{\tau}\;L+\Phi_{A,\sigma}\left(
\frac{\partial L}{\partial\Phi_{A,\tau}}-\partial_{\nu}\left(  \frac{\partial
L}{\partial\left(  \partial_{\nu}\Phi_{A,\tau}\right)  }\right)
+\partial_{\rho}\partial_{\nu}\left(  \frac{\partial L}{\partial\left(
\partial_{\rho}\partial_{\nu}\Phi_{A,\tau}\right)  }\right)  \right)
\nonumber\\
& +\left(  \partial_{\sigma}\Phi_{A,\mu}\right)  \left(  \frac{\partial
L}{\partial\left(  \partial_{\tau}\Phi_{A,\mu}\right)  }-\partial_{\rho
}\left(  \frac{\partial L}{\partial\left(  \partial_{\rho}\partial_{\tau}%
\Phi_{A,\mu}\right)  }\right)  \right)  +\left(  \partial_{\sigma}%
\partial_{\nu}\Phi_{,\mu}\right)  \left(  \frac{\partial L}{\partial\left(
\partial_{\tau}\partial_{\nu}\Phi_{A,\mu}\right)  }\right)  ,\label{EMTENSOR}%
\end{align}
whose conservation $\partial_{\tau}T_{\;\;\sigma}^{\tau}=0\;$can be directly
verified via the equations of motion (\ref{GENEL}).

Next we apply the above general results to our Lagrangian
(\ref{LNL}) together with the realization (\ref{GENSUS}), where
$\Phi_{A}=A_{\alpha}$. The
corresponding derivatives are%
\begin{align}
\frac{\delta L}{\delta A_{\alpha}}  & =0,\;\;\;\frac{\delta L}{\delta
A_{\alpha,\tau}}=-F_{\mu\nu}\chi_{0}^{[\mu\nu][\tau\alpha]}-\frac{1}{2}%
\chi_{1}^{[\tau\alpha]\theta[\gamma\beta]}\partial_{\theta}F_{\gamma\beta
}-\frac{1}{2}\chi_{1}^{[\tau\alpha]\{\sigma\theta\}[\gamma\beta]}\partial_{\sigma
}\partial_{\theta}F_{\gamma\beta},\label{DER01}\\
\frac{\delta L}{\delta\left(  \partial_{\nu}A_{\alpha,\tau}\right)
}  &
=-\frac{1}{2}F_{\theta\sigma}\chi_{1}^{[\theta\sigma]\nu[\tau\alpha]
},\;\;\;\;\;\frac{\delta L}{\delta\left(
\partial_{\rho}\partial_{\nu }A_{\alpha,\tau}\right)
}=-\frac{1}{2}F_{\theta\sigma}\chi_{2}^{[\theta
\sigma]\{\rho\nu\}[\tau\alpha]}.\label{DER23}%
\end{align}
The equations of motion outside the sources can be written as
\begin{equation}
0=\partial_{\tau}H^{\tau\alpha},\label{EQMOUT}%
\end{equation}
where%
\begin{equation}
H^{\tau\alpha}=-\left(  \frac{\delta L}{\delta A_{\alpha,\tau}}\right)
+\partial_{\nu}\left(  \frac{\delta L}{\delta\partial_{\nu}A_{\alpha,\tau}%
}\right)  -\partial_{\nu}\partial_{\rho}\left(  \frac{\delta L}{\delta
\partial_{\nu}\partial_{\rho}A_{\alpha,\tau}}\right)  =\hat{\chi}^{[\tau
\alpha][\theta\psi]}F_{\alpha\beta}.\label{DEFH}%
\end{equation}
Now let us consider the energy-momentum tensor (\ref{EMTENSOR}). Let us
observe that the second term in this equation is precisely $-A_{\alpha,\sigma
}H^{\tau\alpha}$ which is not directly gauge invariant. It can be rewritten as%
\begin{equation}
-A_{\alpha,\sigma}H^{\tau\alpha}=-F_{\sigma\alpha}H^{\tau\alpha}%
-A_{\sigma,\alpha}H^{\tau\alpha}=-F_{\sigma\alpha}H^{\tau\alpha}%
-\partial_{\alpha}\left(  A_{\sigma}H^{\tau\alpha}\right)  ,\label{GINV}%
\end{equation}
by using the equations of motion. The last term is identically conserved and
does not contribute to the corresponding charges. The remaining contributions
are%
\begin{align}
\frac{\partial L}{\partial\left(  \partial_{\tau}A_{\alpha,\mu}\right)
}-\partial_{\rho}\left(  \frac{\partial L}{\partial\left(  \partial_{\rho
}\partial_{\tau}A_{\alpha,\mu}\right)  }\right)    & =-\frac{1}{2}\chi
_{1}^{[\theta\psi]\tau[\mu\alpha]}F_{\theta\psi}+\frac{1}{2}\chi_{2}%
^{[\theta\psi]\{\rho\tau\}[\mu\alpha]}\partial_{\rho}F_{\theta\psi},\nonumber\\
\left(  \partial_{\sigma}\partial_{\nu}\Phi_{,\mu}\right)  \left(
\frac{\partial L}{\partial\left(
\partial_{\tau}\partial_{\nu}\Phi_{A,\mu }\right)  }\right)    &
=-\frac{1}{4}\left(  \partial_{\sigma}\partial_{\nu
}F_{\mu\alpha}\right)
F_{\theta\sigma}\chi_{2}^{[\theta\sigma]\{\nu\tau\}[\mu\alpha ]},
\end{align}
which are naturally gauge invariant. The final gauge invariant, non-symmetric
energy-momentum tensor is
\begin{align}
T_{\sigma}^{\tau}  & =-\delta_{\sigma}^{\tau}\;L-F_{\sigma\alpha}H^{\tau
\alpha}-\frac{1}{4}\left(  \partial_{\sigma}F_{\mu\alpha}\right)  \chi
_{1}^{[\theta\psi]\tau[\mu\alpha]}F_{\theta\psi}\nonumber\\
& +\frac{1}{4}\left(  \partial_{\sigma}F_{\mu\alpha}\right)  \chi_{2}%
^{[\theta\psi]\{\rho\tau\}[\mu\alpha]}\left(
\partial_{\rho}F_{\theta\psi }\right)
-\frac{1}{4}F_{\theta\psi}\chi_{2}^{[\theta\psi]\{\nu\tau\}[\mu\alpha]
}\left(  \partial_{\sigma}\partial_{\nu}F_{\mu\alpha}\right)  .\label{FINEMT}%
\end{align}
A direct but rather long calculation allows to verify the
conservation$\;\partial_{\tau}T_{\sigma}^{\tau}=0$, via the equations of
motion\ (\ref{EQMOUT}).

To express the energy-momentum components in terms of the fields $\mathbf{E}$
and $\mathbf{B}$ we use the following $\left(  3+1\right)  $ splitting%
\begin{align}
W_{\mu\nu}\chi_{1}^{\left[  \mu\nu\right]  \tau\left[  \alpha\beta\right]
}U_{\alpha\beta}  & =4W_{0i}\chi^{\left[  0i\right]  \tau\left[  0m\right]
}_1U_{0m}+2\chi^{\left[  0s\right]  \tau\left[mn\right]}_1\left[  W_{0s}%
U_{mn}-W_{mn}U_{0s}\right]  +W_{ij}\chi^{\left[  ij\right]  \tau\left[
mm\right]}_1 U_{mm},\nonumber\\
W_{\mu\nu}\chi_{2}^{\left[  \mu\nu\right]  \{\tau\rho\}\left[
\alpha \beta\right]  }U_{\alpha\beta}  & =4W_{0i}\chi^{\left[
0i\right]\{\tau \rho\}\left[0m\right]}_2 U_{0m}+2\chi^{\left[
0s\right]\{\tau\rho\}\left[ mn\right]}_2\left[
W_{0s}U_{mn}+W_{mn}U_{0s}\right]  +W_{ij}\chi^{\left[ ij\right]
\{\tau\rho\}\left[mm\right]}_2 U_{mm},
\end{align}
for antisymmetric fields $W_{\mu\nu},\, U_{\alpha\beta}.$ We also
recall the
relation%
\begin{equation}
-L-F_{0\alpha}H^{0\alpha}=\frac{1}{2}\left(  \mathbf{E\cdot D+B\cdot
H}\right)  ,
\end{equation}
which is useful in calculating the energy density.

Let us illustrate the above construction by writing the energy density
$u=T_{0}^{0}\;$and\textbf{\ }the Poynting vector\ $S_{i}=T_{0}^{i}\;$in terms
of the fields $\mathbf{E}$ and $\mathbf{B}$ to first order in\textbf{\ }%
$\tilde{\xi}$\textbf{:}%
\begin{align}
\mathbf{S}  & =\mathbf{E\times B+}\frac{1}{2}\tilde{\xi}\alpha_{10}%
\mathbf{E}\times\partial_{t}\mathbf{E}+\tilde{\xi}\beta_{10}\left[  \frac
{1}{2}\left(  \partial_{t}\mathbf{B}\right)  \mathbf{\times B}-\mathbf{E\times
}\left(  \mathbf{\nabla}\times\mathbf{B}\right)  \right]  \mathbf{+}%
\sigma_{10}\tilde{\xi}\frac{1}{2}\partial_{t}\left[  \mathbf{E}\times
\mathbf{B}\right]  ,\label{POYNTXI}\\
u  & =\frac{1}{2}(\mathbf{E}^{2}+\mathbf{B}^{2})-\frac{1}{2}\beta_{10}%
\tilde{\xi}\mathbf{B}\cdot\mathbf{\nabla}\times\mathbf{B}-\frac{1}{2}%
\alpha_{10}\tilde{\xi}\mathbf{E}\cdot\mathbf{\nabla}\times\mathbf{E}-\frac
{1}{2}\sigma_{10}\tilde{\xi}\mathbf{\nabla}\cdot\left[  \mathbf{E}%
\times\mathbf{B}\right]  .\label{EDENSXI}%
\end{align}
The terms proportional to $\sigma_{10}\;$correspond to the liberty of
modifying the energy momentum tensor\ as%
\begin{equation}
\tilde{T}_{\sigma}^{\tau}=T_{\sigma}^{\tau}+\partial_{\rho}V_{\sigma}%
^{[\tau\rho]},\;\;V_{\sigma}^{[\tau\rho]}=-V_{\sigma}^{[\rho\tau]},\label{emf}%
\end{equation}
which was previously used in Eq.\ (\ref{GINV}). In the $(3+1)$ partition the
above means%
\begin{equation}
\tilde{u}=u-\mathbf{\nabla\cdot Q,\;\;\;\tilde{S}=S}+\partial_{t}%
\mathbf{Q}+\mathbf{\nabla\times W,}%
\end{equation}
with $Q_{i}=V_{0}^{[i0]}$ and $W_{i}=\frac{1}{2}\epsilon_{ijk}V_{0}^{[jk]}%
$. The last terms in Eqs. (\ref{POYNTXI}) and (\ref{EDENSXI})\
correspond to the choice
\begin{equation}
\mathbf{Q}=\frac{1}{2}\sigma_{10}\tilde{\xi}\left(  \mathbf{E}\times
\mathbf{B}\right)  ,\ \ \ \ \ \ \mathbf{W}=\mathbf{0}.
\end{equation}
Thus, the contributions proportional to $\sigma_{10}\;$can be
eliminated and one recovers the corresponding expressions that can
be obtained directly from
Maxwell's equations. Using the fields $\mathbf{E}$\ , $\mathbf{B},%
$\ $\mathbf{D}$\ and $\mathbf{H}$\textbf{, }together with the equations of
motion (\ref{cem}-\ref{ccr}) in vacuum and the freedom given by the
energy-momentum tensor transformation (\ref{emf}), these magnitudes can also
be written to first order in $\tilde{\xi}$\ in a much more compact form as%
\begin{align}
\mathbf{S}  & =\frac{1}{2}\left(  \mathbf{E}\times\mathbf{B}+\mathbf{D}%
\times\mathbf{H}\right)  -\frac{1}{2}\left(  \alpha_{10}-\beta_{10}\right)
{\tilde{\xi}}\left(  \mathbf{B\times}\left(  \mathbf{\nabla}\times
\mathbf{E}\right)  -\mathbf{E\times}\left(  \mathbf{\nabla}\times
\mathbf{B}\right)  \right)  ,\\
u  & =\frac{1}{2}(\mathbf{E\cdot D+B\cdot H}).
\end{align}

\section{Green functions}

The exact retarded Green function for the potential $\mathbf{A}$ in the
circular polarization basis is\ \cite{PDELC,PLBTOBE}%
\begin{equation}
G_{ij}^{ret}(\omega,\mathbf{R})=\int\frac{d^{3}k}{\left(  2\pi\right)  ^{3}%
}e^{i\mathbf{k}\cdot\mathbf{r}}\,\tilde{G}_{ij}^{ret}(\omega,\mathbf{k}%
)=\frac{1}{2}\int\frac{d^{3}k}{\left(  2\pi\right)  ^{3}}e^{i\mathbf{k}%
\cdot\mathbf{R}}\,\sum_{\lambda}G^{\lambda}(\omega,\mathbf{k})\left(
\delta_{ik}-\frac{{k}_{i}{k}_{k}}{k^{2}}+i\lambda\epsilon_{irk}\frac{{k}_{r}%
}{k}\right)  ,\label{gg}%
\end{equation}
where $\mathbf{R}=\mathbf{r}-\mathbf{r}^{\prime},$ ${\hat{k}}_{i}%
=k_{i}/|\mathbf{k}|$, $k=|\mathbf{k}|$, and $G^{\lambda}(\omega,\mathbf{k})$
is obtained from Eq.(\ref{af}),%
\begin{equation}
G^{\lambda}(\omega,\mathbf{k})=\frac{1}{k^{2}Q-\omega^{2}P+\lambda
k\,R},\;\;\;\;\lambda=\pm1.\label{gl}%
\end{equation}
Taking the analytic continuation $\omega\rightarrow\omega+i\epsilon$ to obtain
the causal Green functions, only the poles in the upper half plane
of\textbf{\ }$k$ make a contribution to the integration. By successive
rescalings, the denominator in Eq. (\ref{gl}) can be written in a more
convenient form%
\begin{equation}
Qk^{2}-P\omega^{2}+\lambda kR=-n_{0}^{2}\omega^{2}Qa(M^{\lambda}%
-M_{0})(M^{\lambda}-M_{+})(M^{\lambda}-M_{-}),\label{gl1}%
\end{equation}
where we introduce the notation
\begin{align}
Q &  =h_{1}+d_{3}{\omega\tilde{\xi}},\qquad a=h_{3}n_{0}\chi
, \qquad c=\frac{g}{\omega^{2}n_{0}}\chi, \\
\chi &  =\frac{\tilde{\xi}\omega}{h_{1}+d_{3}{\tilde{\xi}\omega}%
},\qquad M^{\lambda}=\frac{\lambda k}{n_{0}\omega},\qquad n_{0}^{2}=\frac{d_{1}%
}{h_{1}+\omega\tilde{\xi}d_{3}}.\label{xin}%
\end{align}
To study the modifications to the dynamics it is enough to expand each root in
powers of the small parameter $\chi$%
\begin{equation}
M_{0}\simeq\frac{1}{\tilde{\beta}_{1}}\chi^{-1},\ \ \ M_{\pm}\simeq\pm\left[
1+\frac{1}{2}\left(  \tilde{\beta}_{1}-\tilde{\alpha}_{1}\right)  \left(
\lambda\chi+\frac{1}{4}\left(  5\tilde{\beta}_{1}-\,\tilde{\alpha}_{1}\right)
\chi^{2}\right)  \right]  ,\label{P2}%
\end{equation}
where $\tilde{\beta}_{1}=h_{3}n_{0}$ and $\tilde{\alpha}_{1}=g/(\omega
^{2}n_{0})$. Since the parameter $\lambda$ and the momentum $k$ appear only in
the combination $\lambda k$, we have the symmetry property%
\begin{equation}
G^{\lambda}(\omega,k)=G^{-\lambda}(\omega,-k),\label{simprop}%
\end{equation}
which will be useful in the final calculation of the Green functions
$G^{\lambda}(\omega,\mathbf{R})$. In the radiation approximation, the integral
in (\ref{gg})\ produces%
\begin{equation}
G_{ik}^{ret}(\omega,\mathbf{R})=-\frac{i}{\left(  2\pi\right)  ^{2}}\frac
{1}{R}\sum_{\lambda}\frac{1}{2}\left(  \delta_{ik}-n_{i}n_{k}+i\lambda
\epsilon_{ipk}n_{p}\right)  \int_{-\Omega}^{\Omega}kdke^{ikR}\,G^{\lambda
}(\omega,k),
\end{equation}
where $n_{i}=x_{i}/r$ and $R=|\mathbf{r}-\mathbf{r}^{\prime}|.$ From now on we
set $R=r$\ in all denominators and understand that $R=r-n\cdot r^{\prime}\;$in
the exponentials. We also neglect terms of order higher than\textbf{\ }%
$1/r$\textbf{. }The cutoff $\Omega < E_{QG}$ defines the low
energy domain of the effective theory. In this way we identify
$G^{\lambda}(\omega,\mathbf{R})$ as
\begin{equation}
G^{\lambda}(\omega,\mathbf{R})=-\frac{i}{\left(  2\pi\right)  ^{2}}\frac{1}%
{r}\int_{-\Omega}^{\Omega}kdke^{ikR}\,G^{\lambda}(\omega,k).\label{GLAMBDA}%
\end{equation}
The factor $e^{ikR}$ forces us to close the integration contour in
the upper half complex plane, choosing for example a semicircle
with radius $k=\Omega$ , picking up the poles in this region. Our
description is valid only for momenta $k << \Omega$. According to
Eqs. (\ref{P2}), the pole at $M_{0}^{\lambda}$ corresponds to the
momentum value$\;|k_{0}|=\left|  Qh_{3}^{-1}\right|
\,\tilde{\xi}^{-1}.$ In the present\textbf{\
}approximation\textbf{\ }the contribution to the integral of this
pole, together with the one of the semicircle in the upper half
complex plane, can be neglected. The two remaining poles, which
are the ones that contribute to the integral, are located at small
displacements with respect to $|k_{\pm}|=n_{0}\omega << E_{QG}$.
In this way we take%
\begin{equation}
G^{\lambda}(\omega,k)=\frac{1}{n_{0}^{2}\omega^{2}QaM_{0}}\frac{1}{\left(
M^{\lambda}-M_{+}\right)  \left(  M^{\lambda}-M_{-}\right)  }.
\end{equation}
From the leading order expressions in (\ref{P2})\ we conclude that the pole
that contributes in the case $\lambda=+1$ is $\left(  \omega+i\epsilon\right)
n_{0}M_{+}$, while for $\lambda=-1$ it is $-\left(  \omega+i\epsilon\right)
n_{0}M_{-}$. The resulting integral is%
\begin{equation}
G^{\lambda}(\omega,\mathbf{R})=\frac{1}{4\pi Q}\frac{1}{r}\frac{2n_{\lambda}%
}{n_{-}+n_{+}}e^{i\omega n_{\lambda}R},\label{glor}%
\end{equation}
where we have considered that the dominant term in $M_{0}$ yields $aM_{0}=1$.
The refraction indices are%
\begin{equation}
n_{\lambda}(\omega)=\lambda n_{0}M_{\lambda}.\label{REFIND}%
\end{equation}
The minus sign in $n_{-}$ is because $M_{-}$ starts with a $-1$. Up to the
order considered, the refraction indices are given by the expressions
\begin{equation}
n_{\lambda}(\omega)=n_{0}\left[  1+\lambda\left(  \tilde{\beta}_{1}%
-\tilde{\alpha}_{1}\right)  \frac{\chi}{2}+\left(  \tilde{\beta}_{1}%
-\tilde{\alpha}_{1}\right)  \left(  5\tilde{\beta}_{1}-\tilde{\alpha}%
_{1}\right)  \frac{\chi^{2}}{8}\right]  .\label{ri}%
\end{equation}
From Eq. (\ref{ri}), and using Eqs. (\ref{ALPHA}), (\ref{CONSTCOEF}),
(\ref{INHOM1}-\ref{INHOM0}), (\ref{xin}) and (\ref{P2}), we can obtain the
second order expansion for\textbf{\ }$n_{\lambda}$%
\begin{equation}
n_{\lambda}(\omega)\simeq1+\lambda\left(  \tilde{\xi}\omega\right)
n_{1}+\left(  \tilde{\xi}\omega\right)  ^{2}n_{2},
\end{equation}
with the real coefficients $n_{1}$ and $n_{2}$ given by
\begin{equation}
n_{1}=\frac{1}{2}\left(  \alpha_{10}-\beta_{10}\right)  ,\ \ \ \ \ \ n_{2}%
=\frac{1}{8}\left[  \left(  \alpha_{10}-\beta_{10}\right)  \left(  \alpha
_{10}-5\beta_{10}\right)  +4\left(  \beta_{02}-\alpha_{02}\right)  \right]  .
\end{equation}
According to this the phase velocity is not $1$\textbf{. }Due to the
dispersive character of the background it becomes%
\[
v_{ph}(\omega)\simeq1-\lambda\left(  \tilde{\xi}\omega\right)  n_{1}+\left(
\tilde{\xi}\omega\right)  ^{2}\left(  \left(  n_{1}\right)  ^{2}-n_{2}\right)
.
\]
Thus, the Green function in terms of space-time coordinates is%
\begin{align}
G^{\lambda}(\tau,\mathbf{R})  & =\frac{1}{4\pi r}\int_{-\Omega}^{\Omega
}d\omega\frac{2n_{\lambda}}{Q\left(  n_{-}+n_{+}\right)  }e^{i\omega
n_{\lambda}R}e^{-i\omega\tau}\nonumber\\
& =\frac{1}{4\pi r}\int_{-\Omega}^{\Omega}d\omega\left[  1+\frac{\lambda}%
{2}\left(  \tilde{\xi}\omega\right)  \left(  \alpha_{10}-\beta_{10}\right)
-\left(  \tilde{\xi}\omega\right)  ^{2}\left(  \alpha_{20}-\beta_{02}\right)
\right]  e^{i\omega\left[  1+\lambda\left(  \tilde{\xi}\omega\right)
n_{1}+\left(  \tilde{\xi}\omega\right)  ^{2}n_{2}\right]  R}e^{-i\omega\tau}%
\end{align}
where $\tau=t-t^{\prime}$.

If $\;\Omega\rightarrow\infty$ the choice of the poles warrants the causal
behavior of the Green function. But the frequency cutoff could introduce some
violation of causality. To investigate this\textbf{\ }possibility, we compute
the Fourier transform of the Green function to obtain its time dependent
expression, by expanding the integrand in powers of ${\tilde \xi}$%
\begin{align}
G^{\lambda}(\tau,\mathbf{R})  & \simeq\frac{1}{4\pi
r}\int_{-\Omega}^{\Omega }d\omega\left\{  1+\lambda n_{1}\left(
1+i\omega R\right)  \omega{\tilde \xi}-\left[
\alpha_{20}-\beta_{02}-i\left(  n_{2}+n_{1}^{2}\right)  R\omega+\frac{1}%
{2}n_{1}^{2}R^{2}\omega^{2}\right]  \omega^{2}{\tilde
\xi}^{2}\right\}  e^{i\omega
\left(  R-\tau\right)  }\nonumber\\
& =\frac{1}{2\pi r}\left\{  1-i\lambda n_{1}{\tilde \xi}\left(
1+R\partial_{R}\right)
\partial_{R}+{\tilde \xi}^2\left[  \alpha_{20}-\beta_{02}-\left(  n_{2}+n_{1}^{2}\right)
R\partial_{R}-\frac{1}{2}n_{1}^{2}R^{2}\partial_{R}^{2}\right]
\partial _{R}^{2}\right\}  \frac{\sin\left(  R-\tau\right)
\Omega}{R-\tau}.
\end{align}
This shows that the main effect of the cutoff is to spread the
propagating field around the light cone, within a wedge defined by
$R\simeq\tau\pm \pi/2\Omega$. Returning to
$G^{\lambda}(\omega,\mathbf{R})$, Eq. (\ref{glor}), we can
characterize the effect of the cutoff in the causal behavior of
the Green function using the generalized susceptibility theorem
\cite{LANDAU1}, a generalization of the Kramers-Kronig relations.
Its real and imaginary parts
as functions of the frequency $\omega$ are%
\begin{align}
\text{Re}\;G^{\lambda}(\omega,R)  & \simeq\frac{1}{4\pi r}\left\{
\left[ 1+ {\tilde \xi}^2\left(  \alpha_{20}-\beta_{02}-\left(
n_{2}+n_{1}^{2}\right)  R\partial
_{R}-\frac{1}{2}n_{1}^{2}R^{2}\partial_{R}^{2}\right)
\partial_{R}^{2}\right]  \cos\omega R+\lambda n_{1}{\tilde \xi}\left(
1+R\partial_{R}\right)
\partial_{R}\sin\left(  \omega R\right)  \right\}  ,\nonumber\label{re}\\
& \\
\text{Im}\;G^{\lambda}(\omega,R)  & \simeq\frac{1}{4\pi r}\left\{
\left[ 1+{\tilde \xi}^2\left(  \alpha_{20}-\beta_{02}-\left(
n_{2}+n_{1}^{2}\right)  R\partial
_{R}-\frac{1}{2}n_{1}^{2}R^{2}\partial_{R}^{2}\right)
\partial_{R}^{2}\right]  \sin\omega R-\lambda n_{1}{\tilde \xi}\left(
1+R\partial_{R}\right)
\partial_{R}\cos\omega R\right\}  .\nonumber\label{im}\\
&
\end{align}
To have a causal behavior they must satisfy the\textbf{\ }Kramers-Kronig
relation%
\begin{equation}
\left.  \text{Im\ }G(\omega)\right\vert _{KK}=-\frac{1}{\pi}P\int_{-\Omega
}^{\Omega}d\omega^{\prime}\frac{\text{Re}\;G(\omega^{\prime})-\text{Re}%
\;G(\Omega)}{\omega^{\prime}-\omega}%
\end{equation}
which gives%
\begin{align}
\left.  \text{Im}G^\lambda(\omega\text{)}\right\vert _{KK}  & =-\frac{1}{4\pi^{2}%
r}\left\{  \left[  1+{\tilde \xi}^2\left(  \alpha_{20}-\beta_{02}-\left(  n_{2}+n_{1}%
^{2}\right)  R\partial_{R}-\frac{1}{2}n_{1}^{2}R^{2}\partial_{R}^{2}\right)
\partial_{R}^{2}\right]  \ P\int_{-\Omega}^{\Omega}d\omega^{\prime
}\frac{\cos\left(  \omega^{\prime}R\right)  -\cos\left(  \Omega R\right)
}{\omega^{\prime}-\omega}\right.  \nonumber\\
& \left.  +\lambda n_{1}{\tilde \xi}\left(  1+R\partial_{R}\right)
\partial_{R} \
P\int_{-\Omega}^{\Omega}d\omega^{\prime}\frac{\sin\left(
\omega^{\prime }R\right)  -\sin\left(  \Omega R\right)
}{\omega^{\prime}-\omega}\right\}
\label{imgkk}%
\end{align}
For $\omega/\Omega\ll1$ the integrals reduce to%
\begin{align}
 P\int_{-\Omega}^{\Omega}d\omega^{\prime}\frac{\cos\left(
\omega^{\prime }R\right)  -\cos\left(  \Omega R\right)
}{\omega^{\prime}-\omega}  & \simeq2\left(  1-\cos\omega R\right)
\frac{\omega}{\Omega}\cos\left(  \Omega R\right)  -\left[
\pi+\left[  \left(  \Omega R\right)  \left(  \cos\Omega R\right)
-\sin\Omega R\right]  \left(  \frac{\omega}{\Omega}\right)
^{2}\right]  \sin\omega R,\nonumber\\
P\int_{-\Omega}^{\Omega}d\omega^{\prime}\frac{\sin\left(  \omega^{\prime
}R\right)  -\sin\left(  \Omega R\right)  }{\omega^{\prime}-\omega}  &
\simeq2\left(  1-\sin\omega R\right)  \frac{\omega}{\Omega}\cos\left(  \Omega
R\right)  +\left[  \pi+\left[  \left(  \Omega R\right)  \left(  \cos\Omega
R\right)  -\sin\Omega R\right]  \left(  \frac{\omega}{\Omega}\right)
^{2}\right]  \cos\omega R.\nonumber\\
& \label{R77}
\end{align}
Furthermore, in the case of a radiation field $\Omega R\gg1$ and hence the
factors $\cos\Omega R$ and $\sin\Omega R$ become strongly oscillating,
nullifying the contributions of the terms where they appear (which also have a
factor $\left(  \omega/\Omega\right)  ^{n},$with $n\geq1$). Thus we can take%
\begin{equation}
P\int_{-\Omega}^{\Omega}d\omega^{\prime}\frac{\cos\left(
\omega^{\prime }R\right)  -\cos\left(  \Omega R\right)
}{\omega^{\prime}-\omega}\simeq -\pi\sin\omega R,\qquad
P\int_{-\Omega}^{\Omega}d\omega^{\prime}\frac
{\sin\left(  \omega^{\prime}R\right)  -\sin\left(  \Omega R\right)  }%
{\omega^{\prime}-\omega}\simeq\pi\cos\omega R. \ \label{R78}
\end{equation}
 Replacing these integrals in (\ref{imgkk}), we
finally get that if Re $G^\lambda(\omega)$ is given by Eq.
(\ref{re}), the imaginary part of the Green function, Im
$G^\lambda(\omega)$, must be
\begin{equation}
\left.  \text{Im}\;G^\lambda(\omega)\right\vert
_{KK}=\frac{1}{4\pi r}\left\{  \left[ 1+{\tilde \xi}^2\left(
\alpha_{20}-\beta_{02}-\left(  n_{2}+n_{1}^{2}\right)  R\partial
_{R}-\frac{1}{2}n_{1}^{2}R^{2}\partial_{R}^{2}\right)
\partial_{R}^{2}\right]  \sin\omega R-\lambda n_{1}{\tilde \xi}\left(
1+R\partial_{R}\right)
\partial_{R}\cos\omega R\right\}  ,
\end{equation}
which coincides with Eq. (\ref{im}), obtained by direct
computation.

\section{Final remarks}

 To summarize, in the preceding sections we have presented a
general description for a large class of effective models for the
electromagnetic field incorporating  dynamical corrections
motivated by QG  and leading to departures from standard physics.
The main features characterizing the models to which such a
description is applicable are: (1) the validity of gauge
invariance and charge conservation, (2) the use of standard
commuting space-time coordinates together with the corresponding
Fourier transform methods (which is not the case of Double Special
Relativity models, for example), (3) the assumption that effective
field theories constitute an appropriate tool for describing the
low energy behavior of remnant effects which could arise from
quantum gravity, (4) the assumption that low energy dynamics is
linear in the potential field, (5) the inclusion of non-local
effects via the operator character of the generalized
susceptibilities. This description makes it also possible to
include anisotropic corrections in the constitutive relations, for
example via additional non-dynamical tensors arising from
spontaneous Lorentz symmetry breaking, a case which is not
considered in this work.

The proposed formalism is quite similar to the usual
electrodynamics in a medium, except for the non-local character of
the effective QG corrections, mirroring the granularity of the
space-time induced by quantum fluctuations of the metric. This
feature leads to an electrodynamics with non-local constitutive
relations, which contains terms connecting $\mathbf{D}$ and
$\mathbf{H}$ with both $\mathbf{E}$ and $\mathbf{B}$. Thus the QG
modifications can be modelled by a dispersive bianisotropic
medium, where the propagation of the electromagnetic field is
characterized by a refraction index, whose first order term in the
perturbative parameter $\tilde{\xi}$ is directly related to vacuum
birrefringence. We have considered the models from the point of
view of active transformations, i.e. observable Lorentz symmetry
violations associated with boosts in a given reference frame

The effective models correspond in fact to high order theories.
Hence we used an adequate generalization of the Noether theorem to
find the energy-momentum tensor to second order in the LIV
parameter $\hat{\xi}$.  Next we determined the density of energy
and momentum carried by the electromagnetic field, for which we
give explicit expressions to order $\tilde{\xi}$. They acquire a
simple form when written in
terms of the fields $\mathbf{E}$,$\;\mathbf{B}\;$and$\;\mathbf{D}$%
\textbf{,}$\;\mathbf{H}$ , which shows the convenience of the latter for
describing the dynamics, in an analogous way to the usual electrodynamics in media.

This theory is valid only for low energies. We have also studied
the consequences of this fact by using an explicit cutoff
$\;\Omega\ll E_{QG} \sim \tilde{\xi }^{-1}$.  In fact, the results
in Eqs. (\ref{R77}) and (\ref{R78}) show that  the introduction of
the cutoff does not  produce any significant causality violation
in the radiation regime ($\Omega R >>>1$) because the expected
modifications proportional to $\left(\omega/\Omega \right)^n$ in
(\ref{R77}), which are the subject of possible signals of new
physics in these approaches, are further suppressed by highly
oscillating terms proportional to $\sin(\Omega R), \, \cos(\Omega
R)$  thus nullifying the impact of causality violation upon the
corresponding observational effects. The most outstanding
manifestation of the cutoff is a spreading of the propagation of
the electromagnetic field around the light cone. In fact there are
two sources for such a spreading. One is due to the cutoff and the
other arises from the dispersive character of the effective
medium, which leads to an $\omega $-dependent phase velocity. The
relation between both effects depends on the relative value of
$\Omega$ and $\tilde{\xi}^{-1}$. In any case, for distances large
enough from the source ($\omega R\gg1$), the dispersive effect
will finally dominate.

There remains to discuss the causal behavior of the full theory. There are two
possible sources of acausality. One is related to the dispersive character of
the effective medium, while the other is related to the existence of
velocities $v>1$ that leads to photons propagating to the past in highly
boosted reference frames, and hence to the possibility of acausal loops. This
issue is beyond the scope of the present work, and will be discussed in detail elsewhere.

\section*{Acknowledgements}

R.M. acknowledges partial support from CONICET-Argentina. L.F.U is partially
supported by projects CONACYT-40745F,\ CONACYT-47211F and DGAPA-UNAM-IN104503-3.

\end{document}